\documentclass[apj]{emulateapj}

\shorttitle{Foreground Subtraction in 21 cm Measurements}
\shortauthors{Mao}

\begin{document}

\title{Subtracting Foregrounds from Interferometric Measurements of the Redshifted 21 cm Emission}

\author{Xiao-Chun Mao\altaffilmark{1} \altaffilmark{\dag}}

\altaffiltext{1}{National Astronomical Observatories, Chinese Academy of
Sciences, Beijing 100012, China}
\altaffiltext{\dag}{E-mail: xcmao@bao.ac.cn}

\slugcomment{The Astrophysical Journal, 744:29 (9pp), 2012 January 1}
\received{2011 July 6}
\accepted{2011 September 21}
\published{2011 December 9}

\begin{abstract}
The ability to subtract foreground contamination from low-frequency observations is crucial to reveal the underlying 21 cm signal. The traditional line-of-sight methods can deal with the removal of diffuse emission and unresolved point sources, but not bright point sources. In this paper, we introduce a foreground cleaning technique in Fourier space, which allows us to handle all such foregrounds simultaneously and thus sidestep any special treatments to bright point sources. Its feasibility is tested with a simulated data cube for the 21 CentiMeter Array experiment. This data cube includes more realistic models for the 21 cm signal, continuum foregrounds, detector noise and frequency-dependent instrumental response. We find that a combination of two weighting schemes can be used to protect the frequency coherence of foregrounds: the uniform weighting in the \emph{uv} plane and the inverse-variance weighting in the spectral fitting. The visibility spectrum is therefore well approximated by a quartic polynomial along the line of sight. With this method, we demonstrate that the huge foreground contamination can be cleaned out effectively with residuals on the order of $\sim 10$ \rm{mK}, while the spectrally smooth component of the cosmological signal is also removed, bringing about systematic underestimate in the extracted power spectrum primarily on large scales.
\end{abstract}


\section{Introduction}

As a direct probe of the intergalactic medium (IGM), the 21 cm line emitted by neutral hydrogen will provide rather tight constraints on the early phase of cosmic structure formation. Simulations of the IGM evolution have shown that the 21 cm radiation from the epoch of reionization (EoR) has a strength of $\sim10$ mK, and is expected to oscillate significantly with redshift \citep[e.g.][]{Matteo02,Ciardi03,McQuinn06,Jelic08}. Low frequency interferometers like the Low Frequency Array (LOFAR), Giant Meterwave Radio Telescope (GMRT), Murchison Widefield Array (MWA), Precision Array to Probe Epoch of Reionization (PAPER), and 21 CentiMeter Array (21CMA) will aim to seek statistical detections of this cosmological signal in the near future. Unfortunately, the redshifted 21 cm signal is swamped by a long list of contaminants. The presence of Galactic and extragalactic foreground sources, which contribute a brightness temperature on the order of $\sim100$ K at 100 MHz, does pose a serious challenge for the upcoming observations \citep{Shaver99,Furlanetto06}. In this paper, we concentrate on the ability to subtract foregrounds from radio interferometric measurements and further reveal the underlying 21 cm signal.

Over the last decade, much effort has been made in exploring possible methodologies for foreground subtraction \citep[e.g.][]{Matteo02,Oh03,Zaldarriaga04,Furlanetto04,Santos05,Wang06,Morales06,Gleser08,
Bowman09,Liu09a,Liu09b,Harker10,Liu11}. The most widely discussed proposal focused on the line-of-sight (LOS) technique, taking advantage of the foreground smoothness in frequency space. Owing to the ``mode-mixing'' effect, previous studies were confined to the removal of confusion-level contaminants (\emph{i.e.} diffuse emission and unresolved point sources), assuming that the bright and resolved point sources have been cleaned out perfectly by other radio astronomy algorithms such as CLEAN or peeling. However, for the upcoming 21 cm experiments, the subtraction of bright point sources with the required precision is still a problem \citep{Noordam04,Datta09,Datta10,Pindor11,Bernardi11}. On one hand, the deconvolution of point sources is feasible in principle, but introduces some, and perhaps considerable, artifacts due to the lower dynamic range of most radio images. On the other hand, the prior sky models at low frequencies are as yet fairly unconstrained observationally, which need to be improved continually in future measurements.

\begin{figure*}
\begin{center}
\includegraphics[angle=270,scale=0.5]{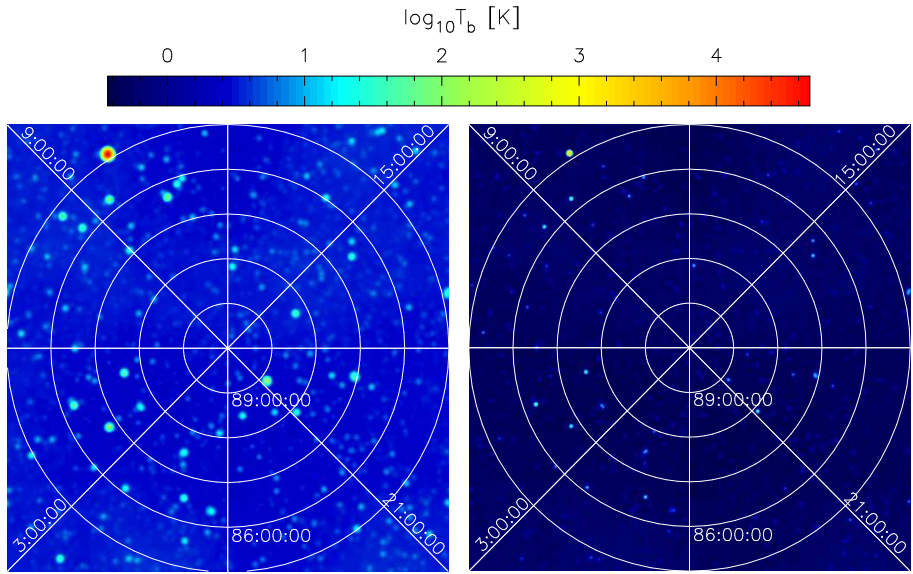}
\caption{Brightness temperature images of the low-frequency foregrounds. The observing frequencies are 80 \rm{MHz} (left panel) and 150 \rm{MHz} (right panel) respectively. Different from previous work, bright point sources are included in our foreground model. The apparent angular size of sources reflects the frequency-dependent size of the synthesized beam in 21CMA observations.}
\end{center}
\end{figure*}

In this paper, we exploit the frequency coherence of continuum foregrounds in Fourier space. Our goal is to develop a blind foreground subtraction technique which allows us to sidestep issues associated with the prior excision of bright point sources. We first simulate the 21 cm interferometric measurements, including the instrumental effects of the frequency-dependent primary beam and \emph{uv} sampling. Techniques are then explored to protect the foreground smoothness along the LOS. With these improved techniques, we show that the visibility spectrum emitted from the resolved and unresolved point sources together with our Galaxy can be well approximated with a smooth function and hence cleaned out simultaneously. While our simulations require specific realizations of the array layout, the proposed method is expected to be applicable for all the first generation EoR experiments. \citet{Zaldarriaga04}, \citet{Liu09b} and \citet{Harker10} have fitted foregrounds as a function of frequency in Fourier space, but they dealt only with the unresolved point sources. \citet{Gleser08} presented a de-contamination approach based on the maximum a-posteriori probability (MAP) formalism, taking into account the bright point sources but not the instrumental response. 

The rest of this paper is organized as follows. In Section 2, we outline the proper simulations of the sky model, and introduce the simulated interferometric measurements. The detailed array parameters that have been used in the simulations are presented here. Our foreground removal technique is described carefully in Section 3. In addition, we measure the impact of foreground subtraction on the cosmological signal, and discuss how the residuals depend on the data reduction and antenna configuration. And in Section 4, we estimate the quality and sensitivity of power spectrum extraction by using the cleaned data cube. The method of suppressing the detector noise is also considered here. Finally, we discuss the implications of the results from our simulations, and present our recommendations for upcoming low-frequency experiments in Section 5.

\section{Sky Model and Radio Interferometry}

\subsection{Simulations of Low-frequency Sky}

In this section, we outline the large-volume simulations adopted in the current work, focusing on the astrophysical foregrounds and the expected 21 cm signal from reionization. All the simulations are performed over a field-of-view of $10^{\circ}\times10^{\circ}$, and the resulting sky maps are arranged onto grids of $500\times500$ pixels. Along the LOS, we use the frequency range extending from 130 MHz to 170 MHz with the resolution of 0.1 MHz.

\begin{figure*}
\begin{center}
\includegraphics[angle=270, scale=0.5]{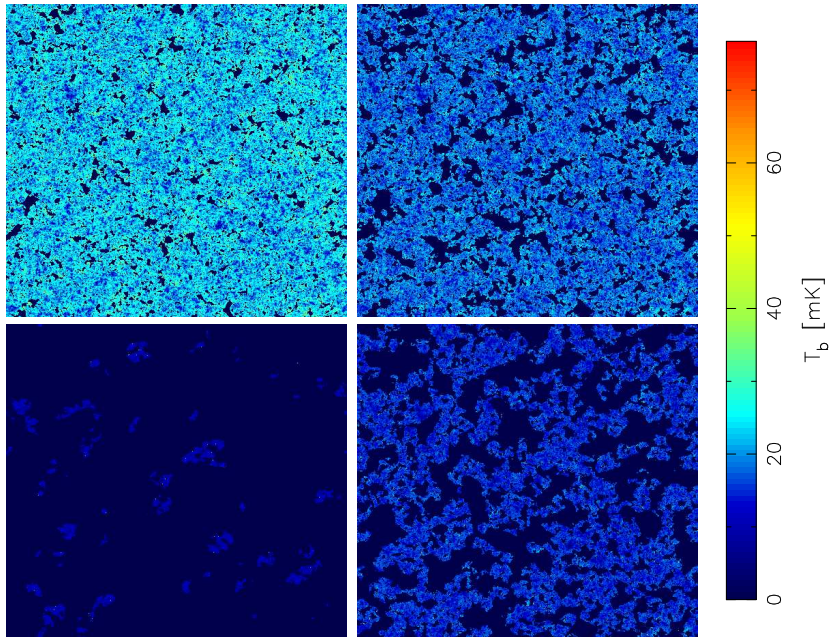}
\caption{Slices through 21 cm brightness temperature box generated from \emph{21cmFAST} simulations, corresponding to $(z,\bar{x}_{\mathrm{HI}})=$ $(10.279,0.753)$,$(9.078,0.573)$,$(8.167,0.342)$,$(6.893,0.024)$ in a clockwise direction.}
\end{center}
\end{figure*}

\subsubsection{Foreground Sources}

To simulate the low-frequency foregrounds with high fidelity, we employ Monte Carlo simulations presented by \citet{Wang10} and references therein, incorporating contributions from three main components: (1)Galactic synchrotron and free-free emission; (2)galaxy clusters; and (3)extragalactic discrete sources such as star-forming galaxies and AGNs. In order to construct the foreground model with higher spatial and spectral accuracies, they first adopt the generic property that radio spectra of foregrounds follow power-law shapes with a running spectral index, and further consider in detail not only random variations of morphological and spectroscopic parameters within the reasonable ranges allowed by multi-frequency observations, but also evolution of radio halos in galaxy clusters, assuming that relativistic electrons are re-accelerated in the intra-cluster medium in merger events and lose energy via both synchrotron emission and inverse Compton scattering with cosmic microwave background (CMB) photons. Our foreground box is kindly provided by the authors.

In some ways, the Galactic radio recombination lines (RRLs) can introduce significant structure in frequency space, but their narrow line widths ($\Delta\nu\sim3$ kHz at 100 MHz) imply that they just occur at very narrow frequency bands. Moreover, \citet{Petrovic10} have proved that the integrated extragalactic radio recombination line background is also unlikely to constitute a formidable foreground. The RRLs are therefore omitted in our analysis, since we can easily excise the contaminated regions of the spectrum in future measurements. The simulated foreground maps as observed by 21CMA are shown in Figure 1. Color versions of the figures are available in the online journal.

\subsubsection{21 cm EoR Signal}

We carry out a publicly available code called \emph{21cmFAST} to generate the expected 21 cm brightness temperature field \citep{Mesinger07,Mesinger10}. For very large volumes, the semi-numerical approach has the advantage of properly and rapidly creating the signals with sufficient resolution. In what follows, we briefly summarize its scheme. The initial conditions in Lagrangian space are initialized at $z=300$. And a Monte Carlo realization of the density field as well as velocity field are then established. Based on this, the non-linear gravitational effects are considered using the first-order perturbation theory as described by the Zel'dovich approximation. In order to increase the speed and dynamic range, the \emph{21cmFAST} does not explicitly resolve source halos. Instead the excursion-set formalism is simply applied to estimate the mean densities around a given point within decreasing sizes, allowing us to obtain the collapsed mass field. The situation and evolution of the ionization fields are directly related to the density distribution. Combined with the velocity gradient and spin temperature, the predicted 21 cm signal from neutral hydrogen can be evaluated through \citep[cf.][]{Furlanetto06}
\begin{eqnarray}
\delta{T_b}(\nu)&=&\frac{T_s-T_{\gamma}}{1+z}(1-e^{-\tau_{\nu_0}}) \nonumber\\
&\approx& 27x_{\rm HI}(1+\delta_{\rm nl}) \Bigg(\frac{H}{d\upsilon_r/dr+H}\Bigg) \Bigg(1-\frac{T_{\gamma}}{T_S}\Bigg) \nonumber\\
&&\times \Bigg(\frac{1+z}{10}\,\frac{0.15}{\Omega_{\rm M}h^2}\Bigg)^{1/2} \Bigg(\frac{\Omega_{\rm b}h^2}{0.023}\Bigg)\, {\rm mK},
\end{eqnarray}
where $d\upsilon_r/dr$ is the comoving velocity gradient along the line of sight. As usually argued, fluctuations in the spin temperature introduce considerable contributions to the EoR signal especially at higher redshifts. Refer to the original papers for more details.

In our work, the $10^{\circ}\times10^{\circ}$ sky region corresponds to a comoving window on the order of $1000\,\mathrm{Mpc}\times1000\,\mathrm{Mpc}$ at EoR redshifts. The initial density field simulation therefore involves a $1000^{3}\,\mathrm{Mpc}^{3}$ cosmological box with $2000^{3}$ cells, \emph{i.e.},$\sim0.5\,\mathrm{Mpc}$ per pixel on a side. To keep a moderate computational time, we choose to smooth the evolved density field and velocity field into a $500^{3}$ grid, and then generate the ionization field with the assumption of $T_s\gg T_{\gamma}$ at lower redshifts.

\begin{figure}
\begin{center}
\includegraphics[angle=270, scale=0.35]{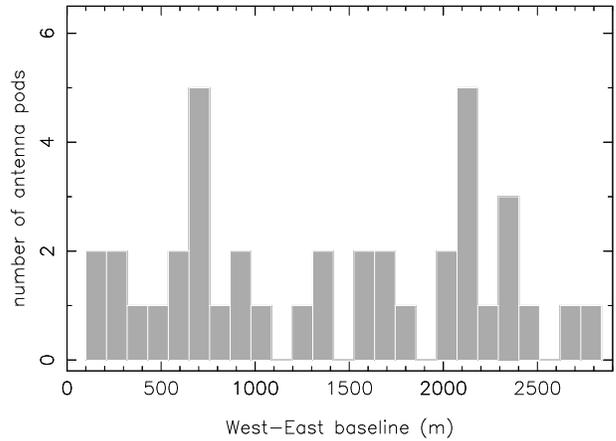}
\caption{Statistic on the locations of the 40 antenna pods distributed from west to east. The number counts are plotted as histograms.}
\end{center}
\end{figure}

Though the 21 cm signal is expected to be at least a factor of $10^4$ smaller than foregrounds, including it in the sky model is necessary for distortion analysis. How does the cleaning process affect the cosmological signal? Since the purpose of this work is not only to develop the foreground removal technique but also to test its usefulness in realistic measurements, a demanding model for the 21 cm signal is important to draw our basic conclusions. In Figure 2, we plot the 21 cm brightness temperature maps at different redshifts from \emph{21cmFAST} simulation boxes.

\subsection{Radio Interferometry with 21CMA}

We employ the West-East baseline of the 21CMA to produce the specific \emph{uv} sampling. There are 40 antenna pods located in this baseline, and the effective collecting area of a pod is 218 $\mathrm{m}^2$ at 150 MHz. All the antennas point toward the North Celestial Pole (NCP), and hence continuously observe a fixed patch in the sky. We show the distribution of the 40 antenna pods in Figure 3. And the integration \emph{uv} coverage is plotted in Figure 4.

For a realistic interferometer, the fundamental observable is a set of complex visibilities, which can be defined as
\begin{eqnarray}
V_{\nu}(u,v)&=&\int\!\!\!\int A_{\nu}(l,m)I_{\nu}(l,m)e^{-2\pi i(ul+vm)}\,dl\,dm
\end{eqnarray}
in the flat-sky approximation. Here, $I_{\nu}(l,m)$ is the sky brightness distribution, and $A_{\nu}(l,m)$ describes the primary beam of an interferometer pair (\emph{i.e.}, normalized reception pattern). For simplicity, we will henceforth use $I'_{\nu}(l,m)$ to denote the modified sky brightness, $A_{\nu}(l,m)I_{\nu}(l,m)$. In practice, the complex visibility can not be known everywhere, but only finite samples are measured on the \emph{uv} plane (as shown in Figure 4). And the sampling process can be described by a \emph{sampling function} $S_{\nu}(u,v)$, which is zero where no data have been taken. As a result, $I_{\nu}(l,m)$ itself can not be recovered directly, instead one obtains the so-called dirty map $I_{\nu}^{D}(l,m)$, where
\begin{eqnarray}
I_{\nu}^{D}(l,m)&=&\int\!\!\!\int S_{\nu}(u,v)V'_{\nu}(u,v)e^{2\pi i(ul+vm)}\,du\,dv,
\end{eqnarray}
and $V'_{\nu}(u,v)$ denotes the noise-corrupted visibilities. Using the convolution theorem for Fourier transform, its relation to the desired intensity distribution $I_{\nu}(l,m)$ can be written as
\begin{eqnarray}
I_{\nu}^{D}(l,m)&=&I'_{\nu}(l,m)\ast B_{\nu}(l,m),
\end{eqnarray}
where the in-line asterisk means convolution, and
\begin{eqnarray}
B_{\nu}(l,m)&=&\int\!\!\!\int S_{\nu}(u,v)e^{2\pi i(ul+vm)}\,du\,dv
\end{eqnarray}
is the synthesized beam or point spread function (PSF) corresponding to the \emph{uv} distribution of baselines. These equations indicate that the measured visibilities in real observations can be simulated as the modified intensity distribution $V_{\nu}$ corrupted by the telescope noise and then multiplied by the \emph{uv} sampling function $S_{\nu}$.

\begin{figure}
\begin{center}
\includegraphics[angle=270, scale=0.4]{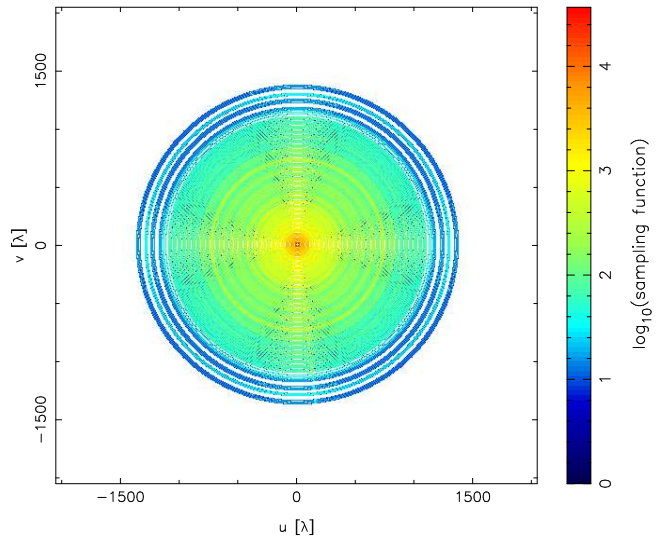}
\caption{Density of visibility measurements in the \emph{uv} plane at 150 \rm{MHz} with integration time of 24 hours. Only the 40 West-East pods of 21CMA are involved.}
\end{center}
\end{figure}

\begin{figure*}
\begin{center}
\includegraphics[angle=270, scale=0.6]{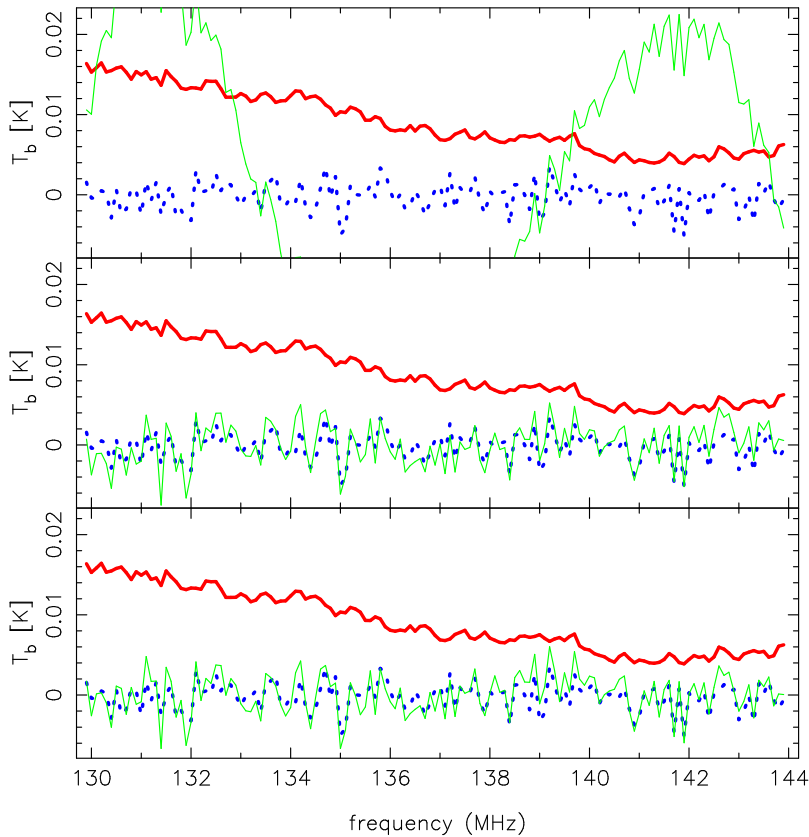}
\caption{Residual spectra after foreground subtraction along the same line of sight. The orders of fitting polynomials are $N=3,4,5$ from top to bottom. The thick solid line and dotted line are the input 21 cm signal and detector noise respectively. And the thin solid line shows the visibility spectrum in the cleaned data cube (after polynomial subtraction), including the residual 21 cm signal, detector noise and fitting errors.}
\end{center}
\end{figure*}

In order to understand how the instrumental response impacts the foreground subtraction, our simulations are passed through the observational pipeline. We first establish the original image cube consisting of the astrophysical foregrounds and the 21 cm signal in our frequency range. And the primary beam can be generally approximated by a Gaussian $A_{\nu}(\vec{\theta})=\rm{exp}(-{\theta}^2/{\theta_b}^2)$ with width $\theta_b \sim 0.6\lambda /D$, where D is the physical size of an antenna pod. At each frequency channel, the sky image is multiplied by the primary beam, and in turn related to visibilities via the two dimensional Fourier transform. Subsequently, we estimate the noise visibilities with one year integration. The rms noise per visibility per frequency channel can be given by \citep{Rohlfs04,McQuinn06}
\begin{eqnarray}
\Delta V_{\nu}^{N}(u,v)=\frac{\lambda^2 T_{sys}}{A_e\Omega_b\sqrt{\Delta \nu t}},
\end{eqnarray}
in which $A_e$ and $\Omega_b$ are the effective area and the beam solid angle of an interferometer element respectively, $\Delta \nu$ is the bandwidth of a single frequency channel, and $t$ is the total integration time for sampling a given $(u,v)$ location. For 21CMA, we assume the sky-dominated system temperature to be $T_{sys} \approx 440[(1+z)/9]^{2.6}$ \rm{K}. Meanwhile, we approximate the integration time $t=\tau N(u,v)$, where $\tau=5$ \rm{s} is the accumulation duration for each visibility measurement, and $N(u,v)$ is the number of independent samples in that pixel \citep{Bharadwaj05,Bowman09}. Because the thermal noise is random, we draw complex visibilities from Gaussian distributions with zero mean and rms described above. From Figure 4, one can infer that the detector noise would increase significantly toward the outer regions of the \emph{uv} plane, owing to the sparse coverage. Finally, the real-world sampling is accomplished carefully in the \emph{uv} plane. The contribution of any single visibility measurement is applied to only one grid cell. We further normalize the baseline distribution to ensure that each pixel in the sampled part of the \emph{uv} plane has the same weight. With the uniform weighting, we artificially emphasize the information contained in long baselines and increase the effective resolution of the derived sky maps. A natural weighting scheme should not be chosen, since the number of visibility measurements in a \emph{uv} pixel changes with wavelength, inducing fluctuations along the frequency direction.

Following all these simulations, we generate our visibility cube representing actual measurements. In the next section, we will concentrate on how to remove the astrophysical foregrounds and reveal the cosmological EoR signal by using the multi-frequency visibilities.

\section{Foreground Subtraction}

As mentioned above, foreground contamination seems formidable in low-frequency experiments, which exceeds the cosmological signal by at least four orders of magnitude. Symmetry differences between the two are therefore well-studied to separate them from each other. Since the EoR emission appears as bumps along both the frequency and angular directions, the redshifted 21 cm signal is expected to be spherically symmetric in 3D space (ignoring redshift space distortions), and fluctuate rapidly in all three dimensions. On the contrary, continuum foregrounds have strong fluctuations in the transverse direction across the sky but weak ones in the radial direction.

In general, a traditional foreground subtraction strategy includes three steps: bright sources removal, spectral fitting, and residual errors subtraction \citep{Morales06}. In order to protect the frequency coherence of foregrounds, bright point sources have to be subtracted down to a 10-100 mJy threshold prior to the LOS spectral fitting step because the incomplete \emph{uv} coverage of interferometer changes with observing frequency, and thus creates different sidelobe patterns across the sky maps, inducing the ``mode-mixing'' effect as emphasized in \citet{Bowman09} and \citet{Liu09a}. We note that since the bright point sources have spectra with the same functional form of the unresolved point sources, the former itself will not destroy the frequency coherence. And the frequency decoherence seen in real space is caused only by the frequency-dependent telescope response. If we can accurately describe the change of the instrumental response, it will not limit the continuum subtraction any more. In Fourier space, one can easily identify pixels with different \emph{uv} sampling, and hence employ an inverse-variance weighting scheme to describe their information content. In this instance, we automatically skip those empty frequency channels at which the points are not sampled, meanwhile, give higher signal-to-noise data points greater weights. Basically, the ``frequency-skipping'' effect protects the foreground smoothness along the LOS. As a result, the first two steps in the traditional strategy can be reduced to one: spectral fitting.

\begin{figure}
\begin{center}
\includegraphics[angle=270, scale=0.4]{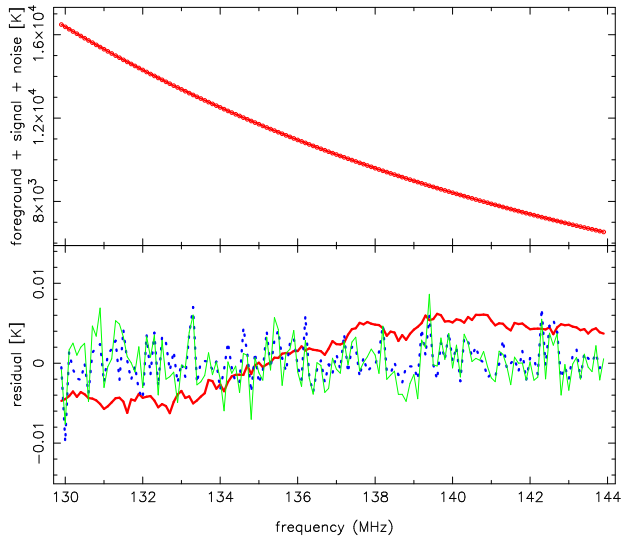}
\caption{Foreground subtraction in a Fourier-space pixel from the part of the \emph{uv} plane where the baseline coverage is complete. Top panel: Frequency spectrum along the line of sight. A foreground fit (solid line) to the visibility data points (open circles) is plotted, with the presence of the 21 cm signal and noise. Bottom panel: Residuals after the fitting polynomial is subtracted from the input data. The thick solid line and dotted line correspond to the original signal and noise, while the thin solid line represents the post-subtraction residuals. Clearly, the spectrally smooth component of the cosmological signal have been unavoidably removed due to the polynomial subtraction. However, fluctuations in the 21 cm signal are preserved well in the residual spectrum.}
\end{center}
\end{figure}

\begin{figure}
\begin{center}
\includegraphics[angle=270, scale=0.4]{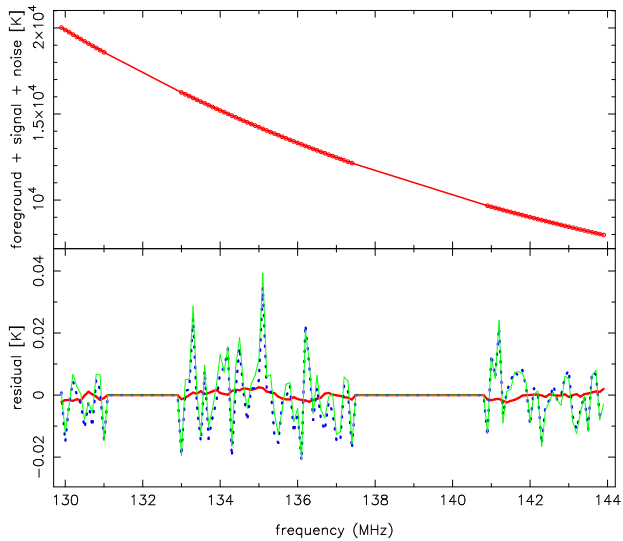}
\caption{Same as Figure 6, but for the polynomial-subtraction technique applied to the part of the \emph{uv} plane where the baseline coverage is sparse. Owing to the sparse \emph{uv} coverage, the detector noise increases significantly and hence dominates the residual visibilities.}
\end{center}
\end{figure}

We now apply our method to the simulated visibility data cube. The frequency range $130\leqslant \nu \leqslant 170$ \rm{MHz} is divided into some sub-bands over which the wavelength varies by less than $10\%$. Foregrounds are then subtracted individually from each sub-band data. This operation offers two major advantages. Firstly, since the bandwidth of sub-band is really small compared to the observing frequency, the primary beams increase slowly toward smaller frequencies and thus the contribution from point sources will be smooth function of frequency that can be accurately matched by the polynomial fit. Secondly, we can expect to extract the HI power spectrum through cross-correlating two sub-bands following which the thermal noise power spectrum do not have to be known. A detailed introduction about this will be given in the next section.

\begin{figure*}
\begin{center}
\includegraphics[angle=270, scale=0.6]{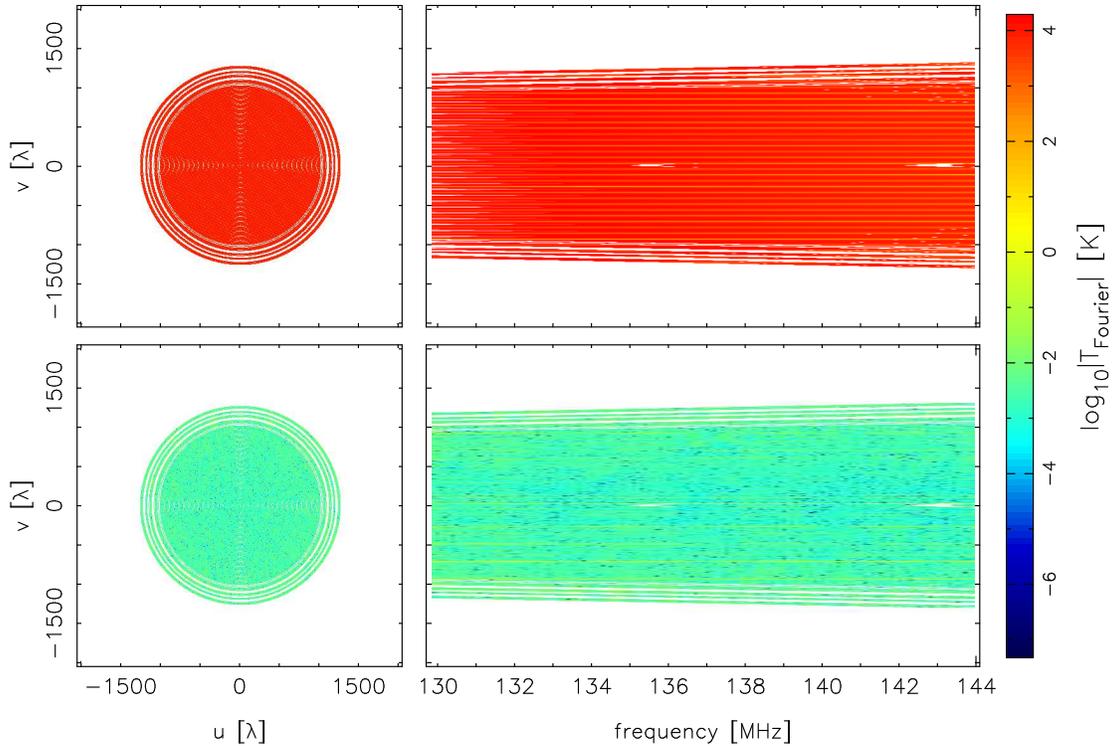}
\caption{Foreground subtraction technique applied to the simulated interferometric measurements. The top row gives two cuts for the input visibility cube, showing a \emph{uv}-map at a single frequency channel $\nu=138$ \rm{MHz} (left panel) and a slice along the frequency direction and \emph{v}-axis in the sub-band (right panel). The bottom row gives the same cuts, but for the cleaned visibility cube following foreground subtraction.}
\end{center}
\end{figure*}

For a given sub-band, there are $m_0$ frequency channels. Within one pixel, we let $y_i$ denote the measured visibility at frequency $\nu_i$ with weight $w_i$, for $i=1,2,\cdots,m_0$. The weight $w_i$ is proportional to the number of baselines that are binned into this \emph{uv} pixel at the frequency $\nu_i$. Because of the ``frequency-skipping'' effect, the effective number of frequency channels used in polynomial fit is reduced to $m$. For our representative method, we directly fit $y_i$ with a set of basis functions
\begin{eqnarray}
\textrm{lg}(y_i)&=&\sum_{k=0}^{N}a_{k}T_{k}(x_i),
\end{eqnarray}
in which $x_i=\textrm{lg}(\nu_i)$, and $T_{k}$ is the Chebyshev polynomials of the first kind. The coefficients $a_{k}$ have the property that they minimize $\sigma$, the sum of squares of the weighted residuals $\epsilon_i$, where
\begin{eqnarray}
\epsilon_i&=&w_i\big[\textrm{lg}(y_i)-f_i\big]
\end{eqnarray}
for $i=1,2,\cdots,m$. Here $f_i$ is the value of the polynomial at the $i$th channels. For the complex visibilities, the real and imaginary parts are fitted separately. Clearly, the key to approximating the visibility spectrum lies in understanding the order $N$ of polynomial. If $N$ is too low, there are insufficient degrees of freedom to remove the foregrounds efficiently; if $N$ is too high, some of the cosmological signal may be mistaken \citep{Furlanetto06}. In fact, the Galactic foreground spectral index fluctuates as a function of both frequency and position. And for extragalactic foregrounds, the sum of power law spectra will not be a power law. Taking these foreground properties and instrumental response into account, we firstly treat $N$ as a free parameter, and fit the visibility spectra with different choices of $N$ along different LOSs. We find that this treatment is not helpful to reveal the faint EoR signal, and introduces structure in the final estimate of power spectrum. In Figure 5, we plot the polynomial-subtraction residuals along the same LOS but for three different orders. One can see that for $N=3$ (top panel), the residual visibility spectrum is dominated by a slowly-varying component, indicating that the foreground contamination is not removed effectively, and for $N=4$ (middle panel) and $N=5$ (bottom panel), the residual spectra have similar shapes. As emphasized by other authors, one should perform the polynomial fit with order as low as possible. Here, the order $N$ is therefore set to be a constant $N=4$ for all pixels.

Figure 6 shows the pristine frequency spectrum (top panel) as well as the post-subtraction residuals (bottom panel) in a \emph{uv} pixel located near the origin. Just as we saw, the huge contaminants (open circles in top panel), including emissions from our Galaxy, galaxy clusters, unresolved point sources and bright point sources, can be well approximated by a smooth function (solid line in top panel) and effectively removed with obviously smaller residuals (thin solid line in bottom panel). The remaining data vary sufficiently rapidly with frequency, which means that the residuals are not dominated by foregrounds. Compared to the original signal and noise (thick solid line and dotted line in bottom panel), one may note that the spectrally smooth component of the cosmological signal has been accidentally removed during foreground subtraction. Fortunately, the effect of foreground cleaning alter the input signal primarily on large scales, and the integral structural component survives from polynomial subtraction. In Figure 7, we present results for a typical \emph{uv} pixel residing in part of the \emph{uv} plane where the baseline coverage is sparse. We can see that it is important to skip the empty frequency channels during the weighted polynomial fit.

In order to protect the spectrally smooth component of the cosmological signal, we try to fit the visibility spectrum with lower frequency resolutions, for $\Delta \nu=1,2,3$ \rm{MHz}: these produced remainders of similar quality. We find that the destruction of large-scale signal would be common to all LOS foreground subtraction schemes.

Figure 8 further illustrates the results of performing the proposed subtraction method on the simulated sky model with frequency-dependent instrumental response. Qualitatively, one can see that the spectral fitting in Fourier space is sufficient to remove the foreground contamination, and the residual visibilities have been suppressed to a level of order $\sim10$ \rm{mK}. This will permit a more simple procedure for foreground cleaning in which the prior removal of bright point sources has been omitted.

Our method is not without disadvantages. In principle, the lower signal-to-noise in each sub-band will degrade the foreground fitting. We have conducted preliminary tests of this method, and found that at least tens of channels are required to give comparable results.

\begin{figure*}
\begin{center}
\includegraphics[angle=270, scale=0.6]{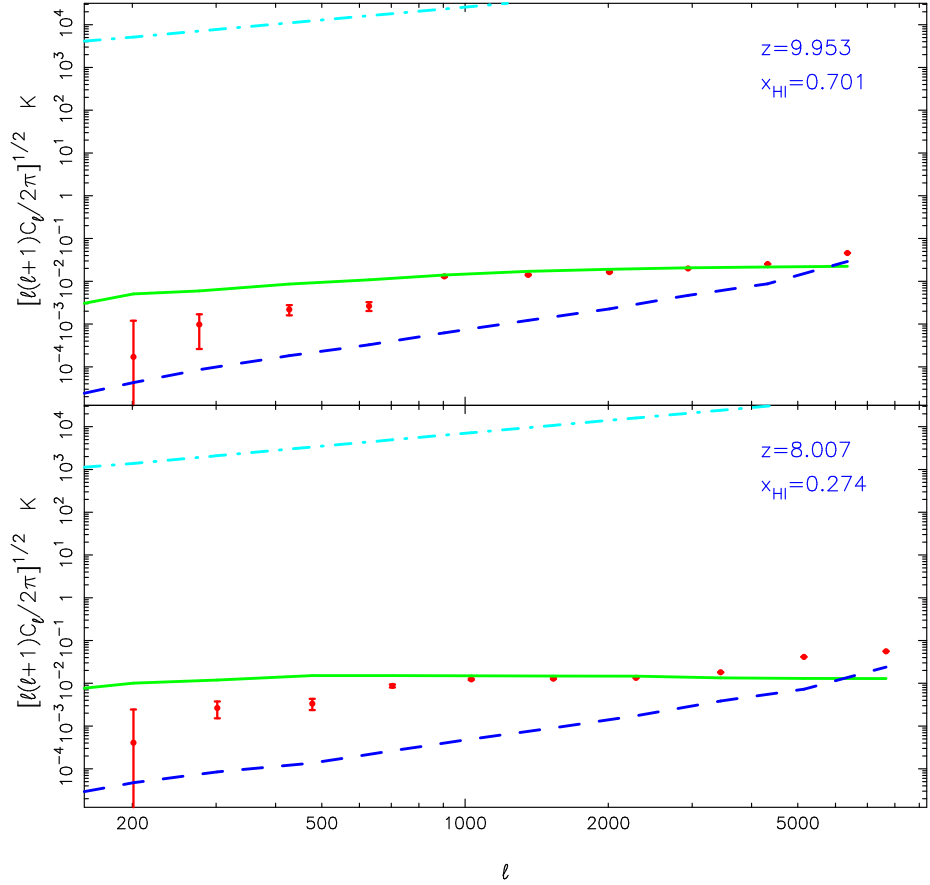}
\caption{Angular power spectra of the input foregrounds (dashed-dotted lines), the 21 cm signal (solid lines), the thermal noise (dashed lines), and the extracted signal (filled circles) at two different redshifts. These power spectra are calculated for one frequency bin of width 0.1 \rm{MHz} from our simulated or cleaned data cubes. The error bars are at the $1\sigma$ confidence level.}
\end{center}
\end{figure*}

\section{Power Spectrum Extraction}

After foreground subtraction, we are left with a residual visibility cube containing 21 cm signal, thermal noise and fitting errors. Since we may be unable to compute the noise power spectrum accurately enough a priori for real experiments, it can not be simply assumed that the desired power spectrum are estimated by subtracting the noise power spectrum from the residual power spectrum. In practice, the possible approach is to explore the independence of thermal noise from two different epochs, and then average their mean power to zero by cross-correlation, leaving only the thermal uncertainty \citep{Bowman09,Harker10}. In addition to the uncorrelation between signal and noise, fitting errors are also uncorrelated with noise during the cross-correlation, and the residual power spectrum therefore contains only three components: signal, fitting errors and their cross-terms. As long as the fitting errors are sufficiently small, this cross-correlation immediately provide us with an estimate of the desired power spectrum.

Using visibility cross-correlation, we define the angular power spectrum \citep{Bharadwaj05,Ali08,Datta10}
\begin{eqnarray}
C_{\ell}=\frac{\sum_{2\pi\mid\mathbf{u}\mid=\ell} W(\mathbf{u})\Big[V^{\ast}_{\nu}(\mathbf{u})V_{\nu+\Delta \nu}(\mathbf{u})+V^{\ast}_{\nu+\Delta \nu}(\mathbf{u})V_{\nu}(\mathbf{u})\Big]}{2\alpha\sum_{2\pi\mid\mathbf{u}\mid=\ell}W(\mathbf{u})},
\end{eqnarray}
where $\mid\!\!\mathbf{u}\!\!\mid=\sqrt{u^2+v^2}$, $\alpha=\pi{\theta_b}^2/2$ describes the effect of primary beam, $W(\mathbf{u})$ and $V_{\nu}(\mathbf{u})$ are the natural weighting function and visibilities measured at frequency $\nu$ respectively. We choose the frequency channels from different sub-bands over which the foreground contaminants are subtracted independently. Two general points are worth noting here. On one hand, the two frequencies $\nu$ and $\nu+\Delta \nu$ are just slightly different, \emph{i.e.} $\Delta \nu/\nu\ll1$. The typical correlation separation of two frequency channels is $\Delta \nu=0.1$ \rm{MHz}. Thus, the frequency separation will cause a very small change in the primary beam, and any evolution of the signal can be neglected in our analysis. On the other hand, accuracy of foreground cleaning diminishes toward the ends of each sub-band, since there are fewer neighboring channels for polynomial fit. As a result, only the central channels of sub-bands should be chosen to estimate the power spectrum. In principle, the cross-correlation will eliminate the thermal noise and preserve the persistent 21 cm signal.

The error bars on the extracted power spectra reflect the statistical errors due to the detector noise as well as sample variance. We calculate the contribution from the noise in a Monte Carlo fashion by measuring the standard deviation of the independent realizations of the thermal noise. And the error from sample variance is estimated by $C_{\ell}/\sqrt{m_{\ell}}$, where $m_{\ell}$ is the number of cells within an annulus near $\ell$. We also confirm the fact that cross-correlation no longer works well when the foregrounds are subtracted over the full bandwidth. In this case, we can not assume the fitting errors and noise are uncorrelated any more.

In Figure 9, we represent the final estimates of angular power spectra at two redshifts. The recovered 21 cm power spectrum (filled circles) appears to be accurate and has small errors at $500<\ell<5000$, while it lies systematically below the input one (solid line) on large scales. The error bars grow in size at large scales due to sample variance. It is clearly that the foreground cleaning process causes suppression of power in the statistical measures. These biases can be correctly accounted for the distortion of the EoR signal, as we see in Fig.6. As pointed out above, the cleaning process removes slowly-varying modes in the simulated data at the expense of attenuating the cosmological signal by accidently removing its large-scale fluctuations. This trend continues as we move to the lower redshift slice (bottom panel). For the sake of comparison, the foreground power (dashed-dotted line) and the thermal noise power (dashed line) before subtraction and cross-correlation are also plotted here. Nearly 1 year is required for the thermal noise to be small so that we could expect the cosmological signal to be the dominant contribution to the angular power spectrum. One can see that it becomes important to include the faint 21 cm signal in the cleaning step, especially for understanding the foreground subtraction technique more comprehensively and for testing its practicability.

\section{Discussion}

In this paper, we studied the foreground removal problem for the upcoming EoR experiments. With simulations of the 21 cm interferometric measurements, we investigated the effect of instrumental response on the foreground removal strategy, and further developed a trend removal technique in Fourier space. Different from previous work, the proposed method treats the confusion-level contaminants and the bright point sources as equivalent, and then clean all such foregrounds simultaneously only using the LOS spectral fitting. We illustrated that this method allows us to avoid complications due to the special treatments to bright point sources, which are unavoidable for traditional foreground subtraction approaches. The basic reason that it works so well is that the frequency dependence of the discrete \emph{uv} sampling can be well described through the inverse-variance weighting scheme in Fourier space.

For our representative method, there is no need to construct proper bright point sources models for cleaning. And its computational cost is in general modest. Moreover, we do not discard any available data, increasing the level of signal-to-noise. Let us mention that the residuals are less than one part in $10^6$ of original sky model, as we see in Fig.8, indicating that the visibilities measured in observations should reach a high dynamic range $>10^6:1$. Furthermore, one should keep it in mind that the required dynamic range of visibility will increase with observed foregrounds. For a field-of-view around NCP (the case of 21CMA), the strength of foregrounds may be little different from our sky model, and thus the dynamic range of measured visibilities should be at least $10^6:1$ to detect the EoR signal.

Although our results are quite encouraging for foreground subtraction, much work remains to be done in this regard. Systematic biases between the input signal and the recovered signal seem to be unavoidable in the LOS polynomial fit. We find that the process of foreground cleaning itself accidentally removes the smooth component of the cosmological signal, and hence leads to suppression in the final estimate of angular power spectrum on large scales. In the current work, these artifacts have been clearly presented. And the astrophysical and cosmological inferences will be significantly distorted, unless the effect of foreground cleaning can be correctly understood. However, for real observations, the systematic underestimate of the true power spectrum would be more difficult to be assessed accurately. Making this sort of correction will always be uncertain, since it depends generally not only on earlier data processing steps, such as instrumental calibration, but also on the statistical properties of the true 21 cm signal. We therefore do not pursue this estimate in the present work. Studying this in more detail in the context of specific experiment must be the subject of future work. And different techniques to remove the foregrounds should also be explored.

Results of this paper further reassure us the astrophysical foregrounds are unlikely to be the main limiting factor in the detection of the EoR information. If the spectrally smooth component is negligible in the true cosmological signal, our proposed method turns out to be feasible. And if not, the systematic biases due to foreground subtraction appear to be the most worrisome remaining problem. Since the proposed method is based on the symmetry differences between foregrounds and cosmic signal, the details of foreground and cosmic models should have no effect on our qualitative conclusions.

\acknowledgments{We thank the anonymous referee for useful comments that have improved this paper. We also thank Xiang-Ping Wu, Max Tegmark, Adrian Liu and Quan Guo for helpful discussions and comments. We are grateful to Jingying Wang, Junhua Gu and Haiguang Xu for providing the foreground data, and to Andrei Mesinger for helpful email correspondence about the 21 cm simulation used here. Support for this work was provided by the National Science Foundation of China (Grant No. 11003019), the Ministry of Science and Technology of China (Grant No. 2009CB824904).}

\end{document}